\documentclass[11pt,a4paper]{article} 
\usepackage{amsmath,amssymb}
\usepackage{epsfig,graphicx,feynmp}

\begin{document}

\title{\bf CHERENKOV RADIATION OF GLUON CURRENTS}
\author{A.~V.~Leonidov$^a$\footnote{Also at the Institute of Theoretical and Experimental Physics.},
M.~N.~Alfimov$^{b}$\footnote{Also at the Moscow Institute of Physics and Technologies.}
\\
$^a$ \small{\em P.N. Lebedev Physical Institute} \\
\small{\em 119991 Leninsky pr. 53, Moscow, Russia}\\
$^b$ \small{\em P.N. Lebedev Physical Institute} \\
\small{\em 119991 Leninsky pr. 53, Moscow, Russia}
}
\date{}
\maketitle

\begin{abstract}
Full quantum calculation of the in-medium Cherenkov gluon emission by gluon current 
in transparent media without dispersion is discussed. 
\end{abstract}

\medskip

One of the natural ways of interpreting experimental data from RHIC \cite{STAR05,STAR10,PHENIX06,PHENIX08a,PHENIX08b} on the double-humped away-side structure of angular correlations is to consider the Cherenkov gluon radiation \cite{D79,D81,ADDK79} as an explanation of this effect \cite{KMW06,MR05,NMR08,DKLV09}.

The phenomenon of Cherenkov radiation has its origin in the nontrivial changes of the dispersion relation for the excitations (quasiparticles) in the medium (as seen from the poles of the propagators):
\begin{equation}
\frac{1}{\omega^2-{\bf k}^2} \Longrightarrow \frac{1}{\omega^2-\epsilon(\omega,{\bf k}) {\bf k}^2},
\end{equation}
where $\epsilon(\omega,{\bf k})$ is a (chromo)permittivity of the medium under consideration. Of special interest are the nonlinear interactions of excitations. The leading nonlinear effect is a three-wave interaction corresponding to the decay of a quasiparticle into two quasiparticles
\begin{equation}
(\omega_1,{\bf k}_1) \to (\omega_2,{\bf k}_2) \oplus  (\omega_3,{\bf k}_3)
\end{equation}
The Cherenkov radiation is a decay of a free vacuum particle into a quasiparticle and a free particle possible for certain special values of 
the permittivity $\epsilon(\omega,{\bf k}) > 1$. In the simplest QED case the Cherenkov radiation is a decay of a free electron into a free in-meidum photon and a free electron \cite{R57}. In this talk we generalize the calculation of \cite{R57} to the three-wave interaction in the in-medium QCD leading to the Cherenkov radiation of gluons by gluon current:
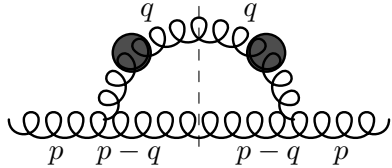
\setlength{\unitlength}{1mm}
\begin{center}
\begin{fmffile}{diagram1}
\begin{fmfgraph*}(50,30)
\fmfleft{i1} \fmfright{o1} \fmf{phantom, label=$p$,
label.dist=10}{i1,v1} \fmf{phantom, tension=1/2}{v1,v2} \fmf{phantom,
label=$p$, label.dist=10}{v2,o1} \fmf{gluon, tension=0.3}{i1,o1}
\fmf{gluon, left, tension=0}{v1,v2} \fmftop{t1} \fmfbottom{b1}
\fmf{dashes, width=.5thin, tension=.5}{t1,v3} \fmf{phantom}{v3,b1}
\fmfiv{label=$q$}{(.4w,.87h)} \fmfiv{label=$q$}{(.6w,.87h)}
\fmfiv{label=$p-q$, label.dist=13}{(.48w,.48h)} \fmfiv{label=$p-q$,
label.dist=13}{(.52w,.48h)} \fmfposition \fmfipath{p}
\fmfiset{p}{vpath(__v1,__v2)} \fmfiv{d.sh=circle, d.f=70}{point
length(p)/4 of p} \fmfiv{d.sh=circle, d.f=70}{point 3length(p)/4 of
p}
\end{fmfgraph*}
\end{fmffile}
\end{center}
From the decay kinematics one immediately sees that the Cherenkov decay can happen only for special angles between the decaying particle and its quasiparticle successor (Cherenkov angle)
\begin{equation}
\cos \theta  = \frac{1}{\sqrt{\epsilon(\omega)}} \left( 1+\frac{\epsilon(\omega)-1}{2} \; \frac{\omega}{E} \right)
\end{equation}
Let us also note that the QFT calculation of this type is possible only for real $\epsilon(\omega )$. In this case it is easy to obtain the following restriction on the energy of the $\omega$ of the Cherenkov gluon:
\begin{equation}
\frac{\omega}{E}\leq \frac{2}{\sqrt{\epsilon(\omega)}+1}
\end{equation}
where $E$ is the energy of the initial decaying gluon.

The Cherenkov gluon emission is of course possible only for special values of energy and momenta of the three participating gluons so that the energy-momentum conservation for the considered decay is fulfilled. To give a quantitative description for this possibility one has to consider an explicit model for the chromopermittivity tensor $\epsilon (\omega,{\bf k})$. Generically chromopermittivity is a nontrivial matrix in the color space  
$\epsilon^{ab} (\omega,{\bf k})$. In what follows we shall confine ourselves to the quasiabelian case, where $\epsilon^{ab} (\omega,{\bf k}) \to \delta^{ab} \epsilon(\omega)$. Close analysis of the experimental data from RHIC suggest the following model for $\epsilon^{ab} (\omega,{\bf k})$ \cite{DKLV09}:
\begin{equation}
 \epsilon(\omega) = \epsilon_0 \cdot \theta (\omega_0-\omega) + 1 \cdot  \theta (\omega-\omega_0) 
\end{equation}
where
\begin{equation}
\epsilon_0 \simeq 5, \;\;\;\; \omega_0 \simeq 3 \; {\rm GeV}
\end{equation}
The Cherenkov decay under consideration is then possible only if $p_0  >\omega_0$, $(p-q)_0  >\omega_0$ and $\omega \equiv q_0 <\omega_0$ and, of course, i.e. we have a decay of the in-vacuum gluon into the in-vacuum and in-medium ones.

The calculation of the energy spectrum of the Cherenkov gluons by gluon current gives \cite{AL10}:
\begin{eqnarray}\label{cherspect}
 P(\omega) & = & 4 \pi \alpha_s N_c \omega  \left(1-\frac{1}{\epsilon} \right) \left(1-\frac{\omega}{E}-\frac{\epsilon-1}{4} \frac{\omega^2}{E^2} \right) \nonumber \\
& \times &  
\left[ 
 1+\frac{1}{2}\left( \epsilon+\frac{\epsilon+1}{1-\frac{\omega}{E}}+\frac{\epsilon}{\left( 1-\frac{\omega}{E} \right)^2} \right)    
 \frac{\omega^2}{E^2}+\frac{(\epsilon+1)^2}{8\left( 1-\frac{\omega}{E} \right)^2}\frac{\omega^4}{E^4} 
\right]
\end{eqnarray}
Let us note that in the soft limit $\omega \to 0$ we get from (\ref{cherspect}) the usual abelian Cherenkov spectrum:
\begin{equation}
 \left. P(\omega) \right \vert_{\omega \to 0} \simeq  
 4 \pi \alpha_s \; N_c \; \omega\left(1-\frac{1}{\epsilon(\omega)} \right)
\end{equation}
The formula (\ref{cherspect}) is the main result we wanted to present in this talk, for more details on this calculation see \cite{AL10}.

The work of A.L. was supported by RFBR grants 09-02-00741 and 08-02-91000-CERN. The work of M.A. was supported by 2010 Dynasty Foundation Grant.

\end{document}